# Coupling of single NV Center to adiabatically tapered optical single mode fiber

Vadim V. Vorobyov[1,2,3], Vladimir V. Soshenko[1,2,4], Stepan V. Bolshedvorskii[1,2,3], Javid Javadzade[3], Nikolay Lebedev[3], Andrey N. Smolyaninov[4], Vadim N. Sorokin[1,2], Alexey V. Akimov[5,2,1]

[1]*Lebedev Physical Institute RAS, 53 Leninskij Prospekt, Moscow, 119991 Russia*
[2]*Russian Quantum Center, 100 Novaya St., Skolkovo, Moscow 143025, Russia*
[3]*Moscow Institute of Physics and Technology, 9 Institutskiy per., Dolgoprudny, Moscow Region, 141700, Russia*
[4]*Photonic Nano-Meta Technologies, Moscow , The territory of Skolkovo Innovation Center, Str. Nobel b.7, 143026 Russia*
[5]*Texas A & M University, Physics Department, 4242 TAMU, College Station, TX, 77843, USA*

e-mail: mrvorobus@gmail.com

We demonstrated a simple and reliable technique of coupling diamond nanocrystal containing NV center to tapered optical fiber. We carefully studied fluorescence of the fiber itself and were able to suppress it to the level lower than single photon emission from the NV center. Single photon statistics was demonstrated at the fiber end as well as up to 3 times improvement in collection efficiency with respect to our confocal microscope.

## 1. Introduction

Applications of solid-state quantum emitters in quantum information processing, quantum communication and sensing are strongly dependent on efficient collection of photons from such an emitter. For example, color centers in diamond, in particular NV center, attracted a lot of attention as a solid-state spin system with long coherence time and unique optical properties [1–10]. Still, collection of emission from this perspective center is a challenging problem. Due to the high refractive index of diamond it is hard to extract fluorescence of color center, with its very broad spectrum creating additional challenges for a cavity based approach [2,11]. Numbers of efforts are made to overcome these challenges. In particular, shaping of nanodiamonds in pillars [12] enables 10 fold improvement in collection of photons with confocal microscope compared to bulk diamond. Another efficient approach enabling improvement of emission collection is implementation of solid immersion lenses, including solid immersion lenses made out of diamond itself [13]. Alternatively, 1D structures have potential to scalability due to directional and confined emission collection from the color center and could be integrated with optical fiber. Such structures utilizing Purcell enhancement can demonstrate considerable improvement in emission collection of nanoscale quantum emitter. For example, plasmonic structures enable quite efficient collection of emission from a single quantum dot or nanodiamond with color center [14–16]. However, this type of structures suffers from substantial losses and is challenging in coupling it with optical fiber due to the mode mismatch. Photonic nanobeams have been demonstrated to be efficiently coupled with tapered optical fibers [17]. Implementation of the diamond nanobeams [18] this way is a very efficient approach, but it still suffers from wide emission spectra of NV center and poor stability of NV's coherence properties to nanofabrication.

Alternative approach to integration of NV center with optical fiber could be direct placement of NV nanodiamond on the waist of pulled optical fiber. Recently atomic force microscope was successfully utilized to place nanodiamond containing NV center onto tapered optical nanofiber and photonic crystal fiber facet [19,20] respectively providing count rate of about 40 kcps in the mode of optical fiber. Here we present a new approach to efficiently couple NV center to the tapered optical fiber and provide detailed analysis of fiber induced background in single photon emission.

## 2. Experimental setup

To analyze emission of NV centers and perform transfer of a nanodiamond between fibers (see below) we used a home built confocal microscope, details of which could be found elsewhere [21]. We used high numerical aperture objective (Nikon 60X PLAN APO NA 0.95) with working distance of 150 micrometers. As an excitation source we used continuous wave laser (Coherent Compass 300, wavelength 532 nm) and as a scanning element we used galvo mirrors (Cambridge Technologies) enabling 100X100 micron field of view for our microscope design (see Figure 1C). This field of view was enough to perform fiber manipulations as described below. We used a combination of notch optical filter centered at 532 and longpass optical filter with cutoff at 600 nm (Semrock, NF03-532E-25, and Semrock LP02-633RU-25) to filter NV's emission from laser pump. In order to identify single NV centers we used time correlation and lifetime measurements. In the first case, we split emission of the NV center into two channels and measured intensity correlation in time domain using time-correlated single photon counting system PicoHarp 300 (Picoquant). To correct on background fluorescence and dark counts of our detectors we implemented noise correction technique while taking second order correlation function. This was done via periodically switching microscope from color center on dark point nearby and recording background from this point. Total counts in each channel were recorded as well. The background corrected second order correlation function $g^{(2)}(\tau)$ could be derived from the definition of the second order correlation function to be:

$$g^{(2)}(\tau) = \hat{g}^{(2)}(\tau) \frac{\langle S_1 \rangle \langle S_2 \rangle + \langle N_1 \rangle \langle S_2 \rangle + \langle N_2 \rangle \langle S_2 \rangle + \langle N_1 \rangle \langle N_2 \rangle}{\langle S_1 \rangle \langle S_2 \rangle} - \frac{\langle N_1 \rangle \langle N_2 \rangle + \langle N_1 \rangle \langle S_2 \rangle + \langle N_2 \rangle \langle S_2 \rangle}{\langle S_1 \rangle \langle S_2 \rangle}$$

Where $\hat{g}^{(2)}(\tau)$ is raw second order correlation function, $\langle S_i \rangle$ is recorded signal in channel $i$ and $\langle N_i \rangle$ is background levels $i$.

For lifetime measurements we used picosecond diode laser LDH-P-FA-530XL (Picoquant) as a laser source with same time-correlated single photon counting system. Due to the fact, that decay of NV center exited state is much longer than the length of the laser pulse fluorescence, recorded after the pulse applied should have single exponential decay. Never the less due to the presence of background fluorescence recorded decay is not a single exponent, but rather combination of fast and slow slope (see Figure 3D). We directly checked by focusing our microscope on bare fiber, that fast component is due to the background, but background has no long tail in decay curve. Therefore, we associated long tail of lifetime with NV center.

The total collection efficiency of our confocal setup was measured to be around 0.5 %.

Our tapering setup is typical for tapering silica-fused optical fibers [22,23]. It consists of two translational stages to pull fiber (Figure 1A); home built heating source on a translational stage; fiber-coupled laser diode; PM100D power meter; computer with the software to control the pulling process. Translational stages in use are Newport M462 stage with a stepper motor coupled to micrometer screw. As a heating source, we use a hydrogen-oxygen flame torch. The gas mixture for the torch is generated from water electrolysis. All mechanics is placed on the 30mm aluminum plate. Relative transmission of the tapered fiber is defined during the process by comparing power transmitted through the fiber to that measured at the beginning of the tapering. For this purpose 100 μW of 637 nm light was coupled into the fiber.

In order to get adiabatically tapered optical fiber we implemented technique similar to that described in [22]. To get a high transmission in the end we took into account that long tapering time leads to fiber contamination while higher taper speed accompanied with higher flame gas speed reduces transmission on the last step. We came to a compromise regime, that allowed us to have up to 90% transmission through the fiber with ~450nm diameter and 5mm length waist (see Figure 1B). Fiber diameter was chosen based on prediction [24,25] with slight correction on broad emission spectrum of NV center. To estimate how broad spectrum will affect optimal fiber diameter we performed calculation for radial dipole located in the center of the fiber using finite-difference time-domain method (FDTD) with open source MEEP [26] software. We found, that in comparison to the single frequency case the maximum of the coupling efficiency shifts towards bigger fiber diameter of 450 nm and has 20% smaller amplitude than in single frequency case. Accordingly to calculation performed at [24,25] this should translate in ¼ NV emission collected by the fiber for radially oriented dipole on the surface and at least factor of 3 less for axial. The last fact helps a lot to reduce the background fiber fluorescence that will be discussed below.

According to our estimates single mode cutoff taken as point at which fraction of the emission collected by the fundamental mode is deviating from 1 was found to be $0.6\ \mu m$, well above chosen fiber diameter. This way chosen fiber diameter should still provide perfect coupling of collected mode to the core-cladding mode at the non-tapered section of the fiber.

## 3. Placing nanodiamond on optical fiber

Our procedure of attaching nanodiamonds to the pulled optical fiber consists of 3 steps. On the first step, we functionalized our nanodiamonds so that their surface became negatively charged. The procedure of functionalization was taken from [27–29]. We started functionalization procedure from nanodiamond solution with concentration of 100 ct/kg. 1 ml of nanodiamond solution was first cleaned in centrifuge and then mixed with 0.1ml of concentrated $HNO_3$ and sonicated for 3 minutes. After that the solution was mixed with 0.9ml of concentrated $H_2SO_4$ and kept at 75°C for 3 days. Then the solution was mixed with water (10 ml) and centrifuged back to original concentration. After that the solution was mixed with 1 ml of 0.4% NaOH solution and kept for 2 hours at 90°C in a water bath. Then the base was replaced with water using centrifuge and the resulting solution was mixed with 1 ml of HCL (concentration 0.4%) heated to 90°C in a water bath and kept for two hours. Finally, acid was replaced with water using centrifuge and the solution was sonicated for 1 hour.

Functionalization does not only change the surface charge of nanodiamond, but also considerably cleans the surface improving nanodiamond photo stability. The second step is the preparation of the donor fiber. The tapered optical fiber was prepared as described above. Then the fiber was placed in Poly-L-Lysine solution (30000-70000 M, Sigma Aldrich), for 15 minutes, while continuously monitoring the transmission, then it was immediately transferred in the mixture of functionalized diamonds with methanol (1:1) for 5 minutes. The transmission of the fiber during this process dropped to the zero due to the scattering by the nanodiamonds. This way the donor fiber was coated with relatively high concentration of nanodiamonds, still enabling us to select single color centers with confocal microscope (Figure 2C).

The last step is the nanodiamond transfer from donor to acceptor fiber. As an acceptor fiber, we used tapered optical fiber prepared as described above but without any chemical coating. Donor and acceptor fiber were crossed under the confocal microscope, without mechanical contact between them. In order to visualize tapered region of the donor fiber we used scattering on nanodiamonds of light coupled to the fiber. For this purpose, we used the same laser source as for fiber transmission. This scattered light greatly help quickly bring two fibers close to each other without making contact between them. Then a single NV center on the donor fiber was located by terms of lifetime measurement followed by investigation of the second order correlation function. The acceptor fiber was moved under the microscope to the location of single NV center and is brought into contact with the donor one. In our setup, acceptor fiber was oriented horizontal with respect to the optical table, while donor fiber was perpendicular to the surface. This was possible due to the horizontal orientation of objective in our microscope. To be consistent we always placed point of contact of two fiber above the nanocrystal we selected for transfer. Then the acceptor fiber was moved downwards (Figure 2A,C). Even though nanodiamond is tiny, ones acceptor fiber touches it, nanodiamond create considerable mechanical contact. While motion in continued fiber follow together until easily observable under microscope jump of the fiber, indicating, that mechanical contact was broken. Usually we had to move not more than 2 µm until jump occur. The jump indicated that nanodiamond with probability of 85% (17 experiments from 20 were successful in transferring desirable NV nanocrystal) was transferred on the acceptor, only in 15% of the cases nanodiamond remain on donor after jump was observed. After transfer procedure the donor was brought out of focus and the acceptor was inspected under the confocal microscope to confirm successful transfer of single nanodiamond with color center (Figure 2B,D).

## 4. Results

In order to prove that the nanodiamond transferred have the only one NV center in it we measured second order correlation function $g_2(\tau)$ with our confocal microscope (Figure 3A). In order to maximize signal collected with microscope or optical fiber we usually selected excitation slightly above saturation for measurement of second order correlation function (see Figure 3C). Measurement of level of counts collected from single NV center in saturation also enable us to compare collection via objective and via optical fiber. It is important to notice, that power used to excite NV center as well as polarization of excitation light generally was different before and after transfer. This was done in order to minimize fiber fluorescence as described below. Also, during the transfer process orientation of NV axis may change a bit causing different excitation efficiency. Therefore shape of the second order correlation function is different on Figure 3 A and B. It is well known, that width of second order correlation function is function of power (width linearly decrease with power) and level of wings of the function (above unitary for NV center) strongly depend on level structure and power applied [30]. Key properties of NV center are related to the fact, that NV center is not 2-level system with dark state and complicated ionization dynamics [31]. Therefore shape of $g_2(\tau)$ is strongly dependent on excitation power [32,33]. But, at the same time level of second order correlation function at zero delay is independent on excitation and may only change with power due to time resolution or noise limitations. The latter option is almost completely removed in our case due to noise compensation procedure described above. Moreover, level of second order correlation function at zero delay is crucially sensitive to number of NV centers. If during transfer process two NV centers will be transferred, then $g_2(0)$ level of will be larger than ½. Therefore the fact, that $g_2(0)$ is not changed during transfer providing single photon statistic is quite convincing argument of transfer success. The other important argument, confirming that exactly the NV we targeted was transferred in measurement of the NV lifetime. Typical distribution of NV center lifetime in nanocrystals we studied has width of 10 ns with average around 18. Despite of this huge spread we measured lifetime of transferred nanocrystal to be within 1 ns on donor and acceptor fiber. Example of such a measurement is presented on (Figure 3D). Here lifetime of NV center was found to be 21 ns on both fibers and confirms that we see the same NV center as before the transfer. It is important to mention, that lifetime of the nanocrystal is not only function of nanocrystal itself, but also of the substrate it is on [33], but since in our case both substrates are similar tapered optical fiber we expect minor difference in the measured lifetime.

Important parameter characterizing the efficiency of interface proposed is the number of counts we collect via optical fiber. Unfortunately optical fiber under green excitation has Raman scattering and it's own fluorescence. While Raman could be cut to high degree with longpass filters, the Thorlabs SM600 optical fiber has fluorescence spectrum considerably overlapping with the one of NV center (Figure 4A). This fluorescence has also lifetime much longer, than NV fluorescence lifetime, reaching value of $30\ \mu s$ (Figure 4B) and thus making time resolved measurement useless for separation single photon and intrinsic fiber emission. Therefore, it is extremely important to minimize contribution of fiber fluorescence. This long lifetime and spectra indicate that florescence of the fiber is most likely caused by oxygen color centers in silica [34].

We believe that the key process causing the fiber fluorescence is that green light scattered on nanodiamond couples into the fiber. This was verified by focusing of the green light on the tapered optical fiber before the nanodiamond transfer and also in location close to the transferred nanoparticle. In both cases no florescence was observed at the fiber end. To minimize effect of the green light coupling we aligned its polarization along the fiber. NV centers we select for transfer will be best of all seen if axis of NV center is oriented along objective axis. For such NV center it's emission would not be dependent on excitation orientation (with minor correction on polarization sensitivity of the microscope which was measured independently), therefore only fiber fluorescence will be affected by polarization of excitation. Nevertheless natural bias toward axis of NV center along objective axes in not absolute and suppression achieved with polarization varies from nanocrystal to nanocrystal. Maximum suppression we observed using polarization of excitation light was factor of 2 with average around 1.5 Figure 4C).

Further suppression of the fiber fluorescence could be achieved using photo bleaching effect known in optical fibers [34]. After sending 30 μW of the green laser emission into optical fiber for 60 seconds the fluorescence of the fiber dropped by the factor of 2 and then slowly recovered over several minutes (Figure 4D). This gave enough time to measure single photon statistics at the fiber output. The results of such measurements are presented in Figure 3B. Here cross correlation between fiber output and the emission collected with the objective was measured. Again, single photon statistics is clearly seen.

The maximum ratio of photons collected via fiber in single photon regime was 3 while on average for all nanocrystal we achieved factor of 1.8. Taking factor of 3 we can calculate total collection efficiency of the fiber as 1.5% from NV center emission in one side of the fiber or 3% in two sides. This number is not limited by NV – fiber coupling but rather by quality of hand–made fiber connections that limited fiber to detector transmission on the level of 10 %. Taking into account these losses we can conclude, that collection at tapered section collection was about 15% (30% two sides), which is a bit lower than predicted numerically. The discrepancy is likely to be due to the fact that NV center has finite size and is placed outside of the fiber [24]. We expect that above-mentioned level of collection could be achieved from the single NV center with manufactured fiber connectors.

## 5. Conclusion

We have demonstrated a robust experimental technique of attaching single nanodiamond containing NV center to the waist of tapered optical single mode fiber. By careful studying of the fiber emission, we were able to suppress it by factor of 4 this way enabling observation of single photon emission at the fiber end. We also demonstrated up to 3X enhancement of the single photon emission with respect to our confocal setup.

## 6. Acknowledgments

We thank Professor Alexey Rubtsov and his group for providing us access to computational resources. This work was supported by RFBR grant # 14-29-07127.

# 8. Figures

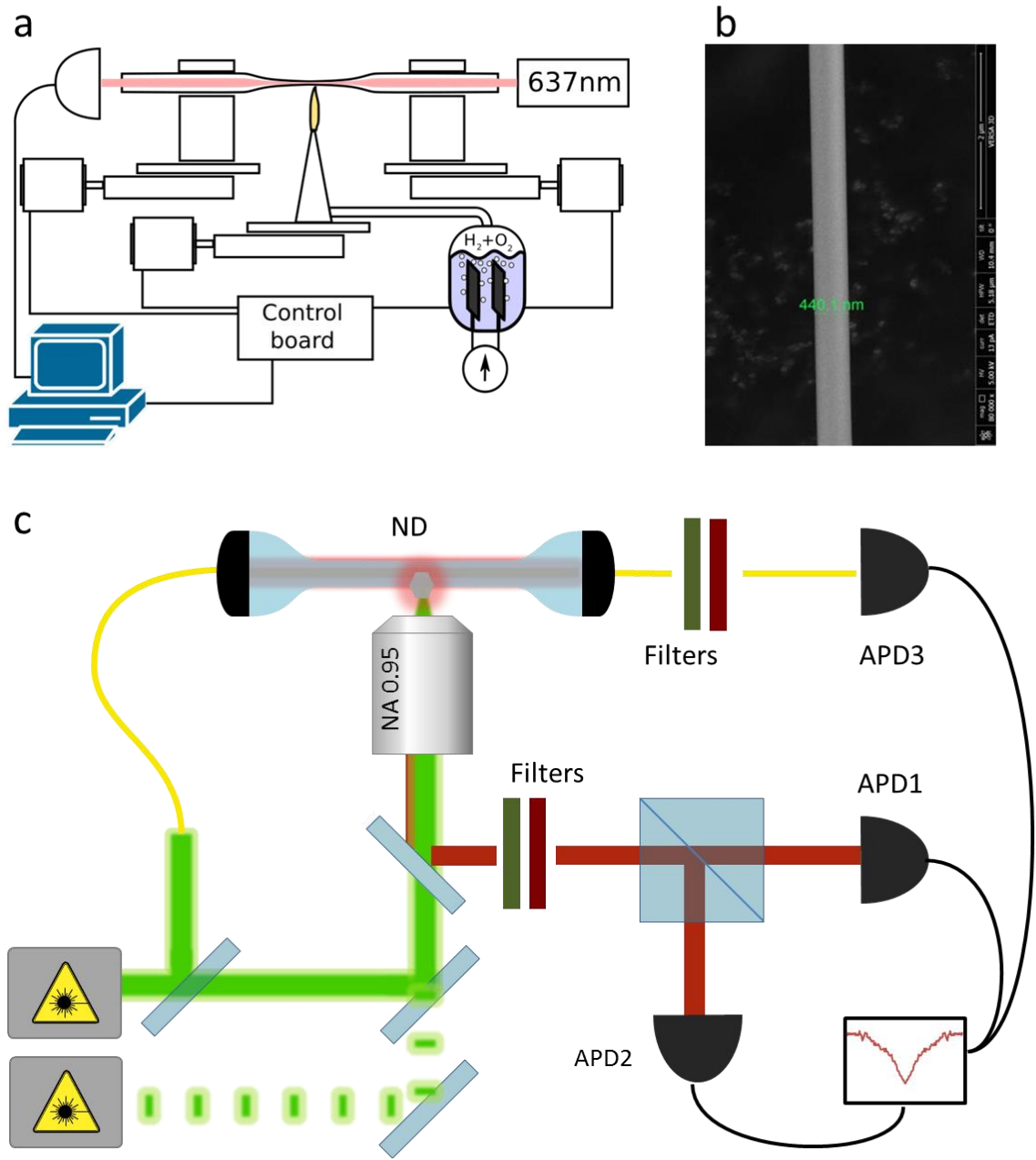

Figure 1 Experimental setup. A – setup for fiber pulling B – Tapered optical optical fiber, C – home-made confocal microscope.

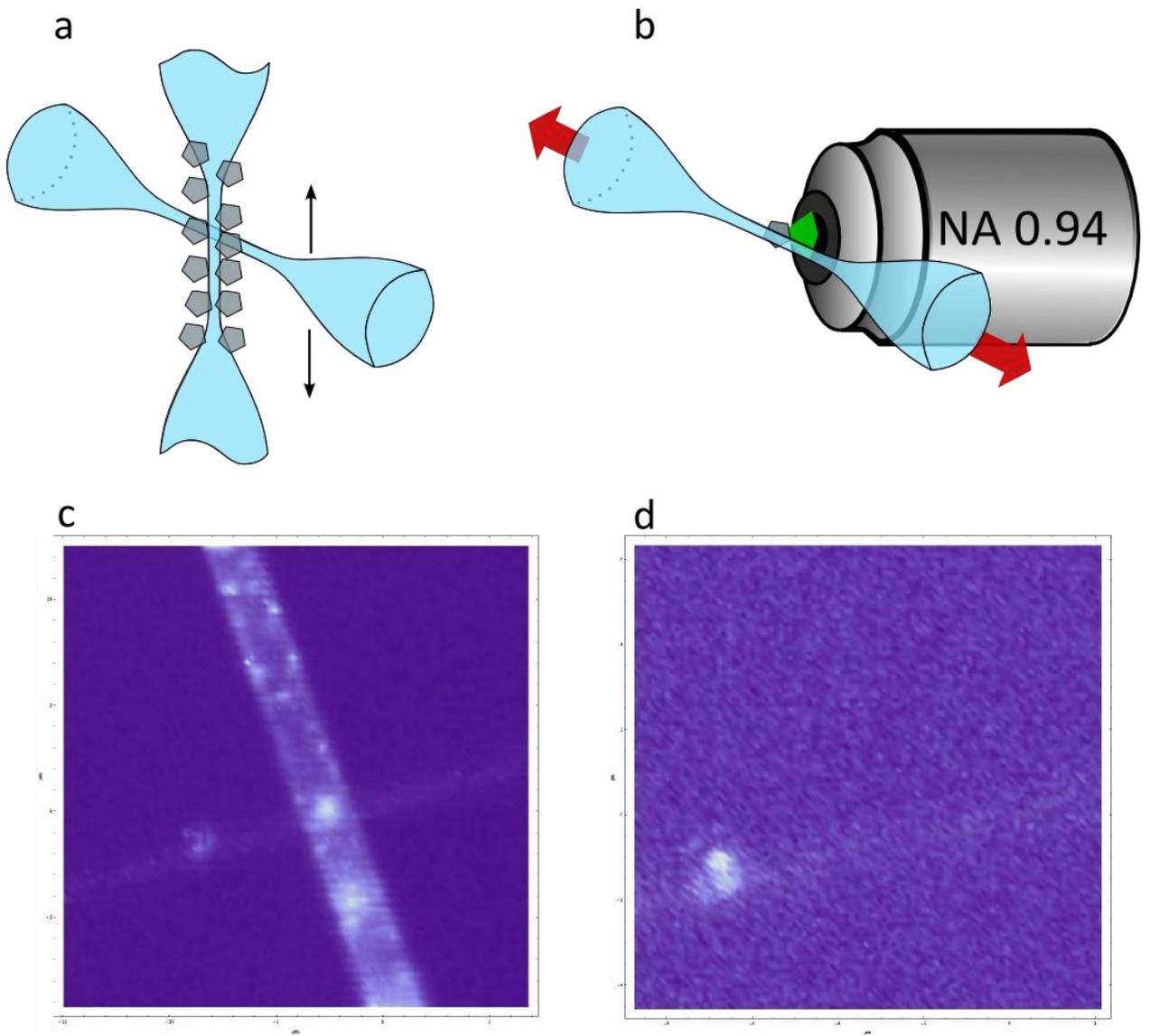

*Figure 2 Attaching nanodiamond to the tapered optical fiber. A) Schematic of the transfer process. Vertical fiber is the donor one, horizontal is the acceptor. B) Inspection of the acceptor fiber under confocal microscope. C) Image of donor fiber (vertical) crossed with acceptor fiber under the confocal microscope. D) Single transferred NV center on the acceptor fiber.*

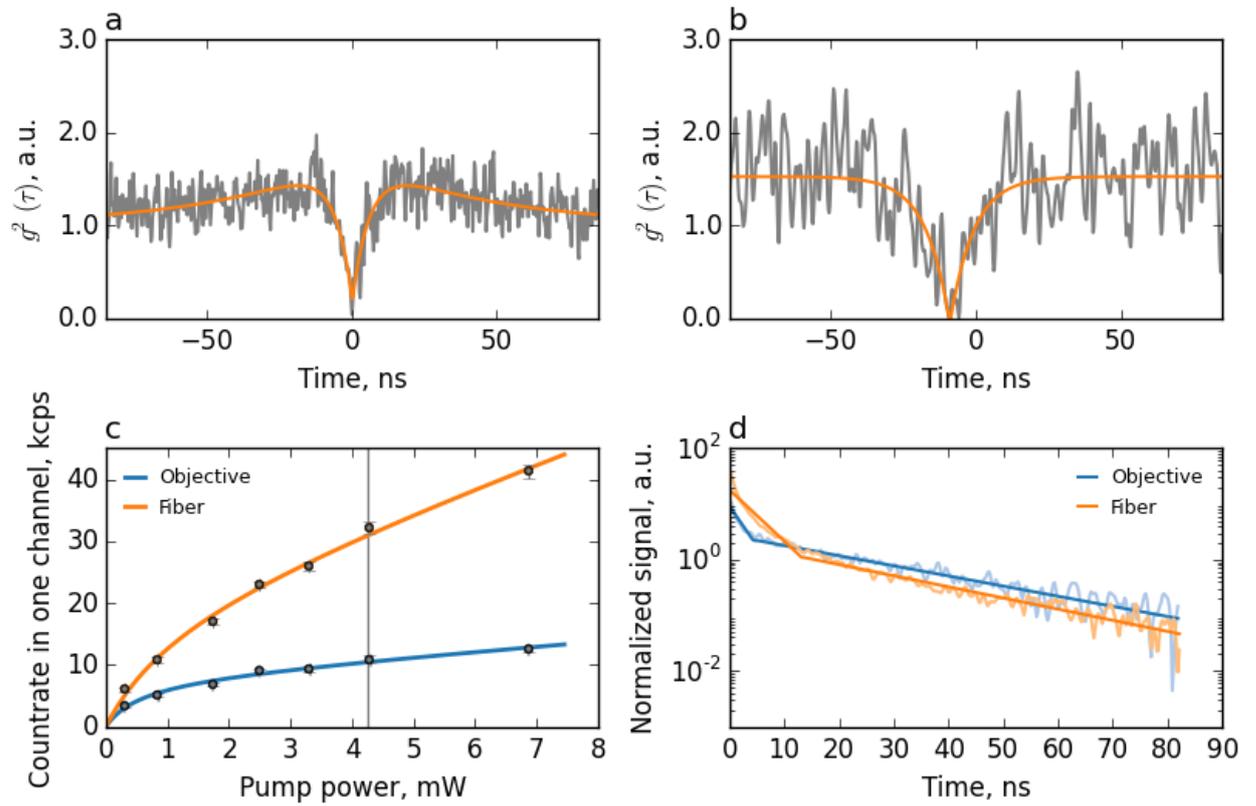

*Figure 3 Experimental results. A) Second order correlation function of selected color center with observation via confocal microscope. B) Second order cross correlation function of selected color center with observation via fiber and confocal microscope, C) number of counts collected via optical fiber in comparison to the confocal microscope for different laser powers. Gray vertical line corresponds to the power at which figure b was taken. D) lifetime of the NV center measured via confocal microscope and via optical fiber.*

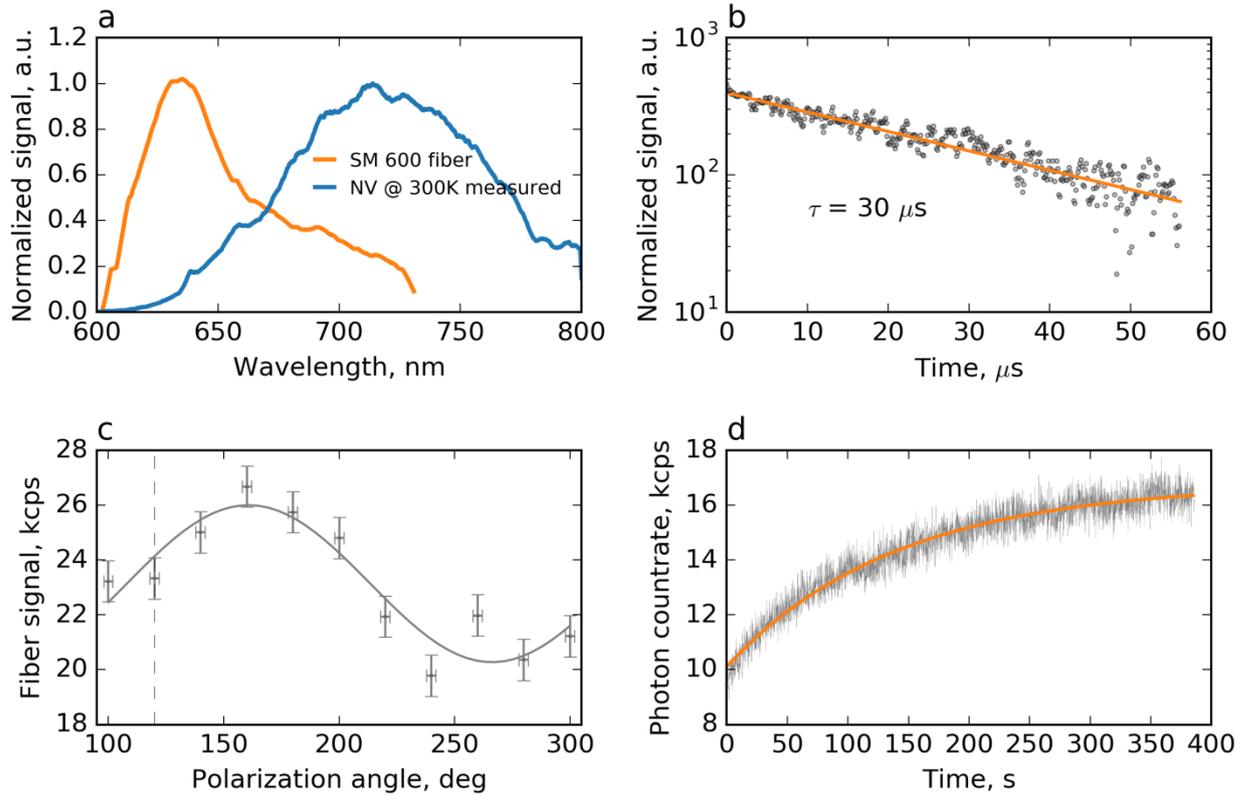

*Figure 4 Optimization of fiber output signal to noise ratio. A) Spectra of fiber fluorescence versus NV fluorescence. Measurements of NV center spectrum was done on NV ensemble. B) Lifetime measurement of fiber fluorescence performed on non-tapered optical fiber. C) Typical polarization dependence of total signal collected via optical fiber. Maximum of the curve correspond to the maximum contribution of fiber fluorescence. Gray vertical dashed line corresponds to the point at which single photon statistic was taken. Solid line represent fit by cosine dependence. D) Recovery of the tapered optical fiber fluorescence after 30s photo bleaching.*